\documentclass[%
 reprint,
superscriptaddress,
 amsmath,amssymb,
 aps,
prx,
]{revtex4-1}

\usepackage{xcolor}

\usepackage{braket}
\usepackage{graphicx}
\usepackage{dcolumn}
\usepackage{bm}

\begin{document}

\preprint{APS/123-QED}

\title{Single-spin quantum sensing: A molecule-on-tip approach}

\author{A. F\'{e}tida}
\author{O. Bengone}
\author{M. Romeo}
\author{F. Scheurer}
\affiliation{Universit\'{e} de Strasbourg, CNRS, IPCMS, UMR 7504, F-67000 Strasbourg, France}
\author{R. Robles}
\affiliation{Centro de Física de Materiales CFM/MPC (CSIC-UPV/EHU), Paseo Manuel de Lardizabal 5, 20018 Donostia-San Sebasti\'an, Spain}
\author{N. Lorente}
\affiliation{Centro de Física de Materiales CFM/MPC (CSIC-UPV/EHU), Paseo Manuel de Lardizabal 5, 20018 Donostia-San Sebasti\'an, Spain}
\affiliation{Donostia International Physics Center (DIPC), 20018 Donostia-San Sebasti\'{a}n, Spain}
\author{L. Limot}
\email{limot@ipcms.unistra.fr}
\affiliation{Universit\'{e} de Strasbourg, CNRS, IPCMS, UMR 7504, F-67000 Strasbourg, France}

\begin{abstract}
Quantum sensing is a key component of quantum technology, enabling highly sensitive magnetometry. We combined a nickelocene molecule with scanning tunneling microscopy to perform versatile spin sensing of magnetic surfaces, namely of model Co islands on Cu(111) of different thickness. We demonstrate that atomic-scale sensitivity to spin polarization and orientation is possible due to direct exchange coupling between the Nc-tip and the Co surfaces. We find that magnetic exchange maps lead to unique signatures, which are well described by computed spin density maps. These advancements improve our ability to probe magnetic properties at the atomic level.
\end{abstract}
\date{\today}


 \maketitle
Quantum sensors can achieve sensitive magnetometry through changes in their spin states or quantum coherence when they interact with a magnetic target. Examples include SQUID magnetometers employing nitrogen vacancy centers in diamond \cite{Ganzhorn2013} or carbon nanotubes \cite{Thiel2016}. The inclusion of magnetic field gradients into magnetometers through scanning probe techniques has allowed for the detection of single spins with a remarkable resolution of up to 10 nm \cite{Rugar2004,Balasubramanian2008,Grinolds2011}. Nevertheless, a paradigm shift is needed to achieve the ultimate aim of atomic-scale resolution in magnetometry.

\begin{figure}
  \includegraphics[width=\columnwidth]{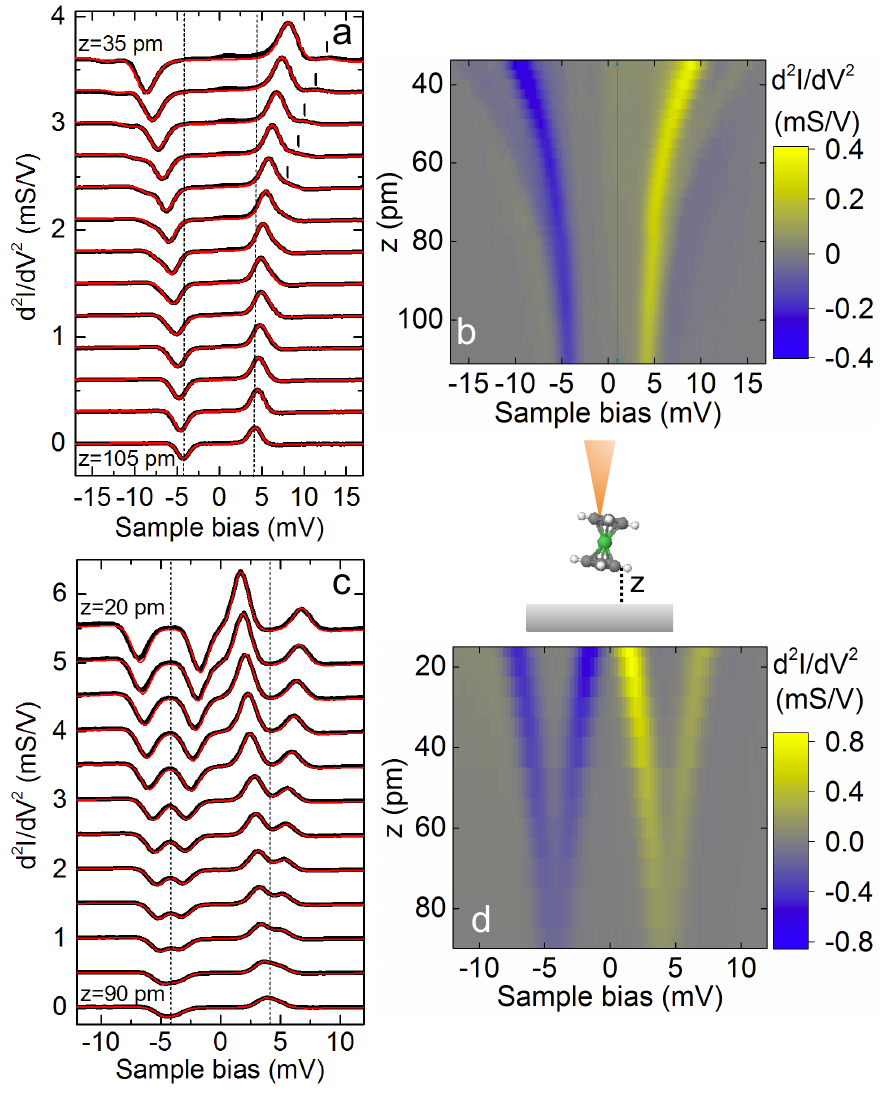}
  \caption{$d^2I/dV^2$ spectra above a Co atom of a (a) monolayer and (c) trilayer island acquired at different tip-Co distances. The spectra are displaced vertically by $0.5$ ms/V from one another. The vertical lines in panel (a) indicate the position of excitation $\ket{\Psi_0}\rightarrow \ket{\Psi_2}$. Red lines: Simulations based on the dynamical scattering model. 2D intensity plot of a series of distance-dependent $d^2I/dV^2$ spectra above a Co atom of a (b) monolayer and (d) trilayer island. Inset: Sketch of the tunnel junction. Nickelocene comprises a single Ni atom sandwiched between two cyclopentadienyl rings (C$_5$H$_5$).
\label{F2}}
\end{figure}

A possible approach involves using a single magnetic molecule as a spin sensor. Molecules offer the advantage of a tunable electronic structure, making them well-suited for fulfilling the design requirements of a quantum sensor \cite{Yu2021}. Notably, recent advancements in scanning tunneling microscopy (STM) techniques have enabled the detection of a molecular spin on a surface through electron spin resonance (ESR) \cite{Willke2021,Kawaguchi2023}. This breakthrough enables the detection of exchange and dipolar fields ($>10$ mT) in the immediate vicinity of the molecular environment \cite{Zhang2022,Zhang2023}. To freely access all surface locations, a more advantageous strategy involves attaching the molecule to the STM tip apex. To date, this has been successfully demonstrated solely with the spin-$S=1$ nickelocene molecule [Ni(C$_5$H$_5$)$_2$, see Fig.~\ref{F2}]~\cite{Ormaza2017b}, albeit at the cost of limiting sensitivity to exchange fields $>1$ T \cite{Czap2019,Verlhac2019}.

In contrast to ESR, the spin states of a nickelocene tip (referred to as the Nc-tip) are in fact monitored through spin excitations arising from the inelastic component of the tunneling current \cite{Heinrich2004}. Sub-angstrom precision in sample spin-sensing is made possible by the exchange interaction occurring across vacuum between the Nc-tip and the surface, which modifies the Nc spin states \cite{Czap2019,Verlhac2019,Wäckerlin2022}. In this study, we examined prototypical ferromagnetic Co islands that were grown on a pristine Cu(111) surface, including monolayer-thick islands that remained unexplored. Our findings, supported by \textit{ab initio} calculations, reveal that the thickness of the ferromagnetic layer exerts control over the orientation of sample magnetization and spin polarization. Furthermore, we demonstrate that the spatial dependence of the exchange energy across the surface is well captured by computed spin-density maps. Our study demonstrates the effectiveness of a Nc-tip in probing surface magnetism, even in the absence of an external magnetic field.

\begin{figure}
  \includegraphics[width=\columnwidth]{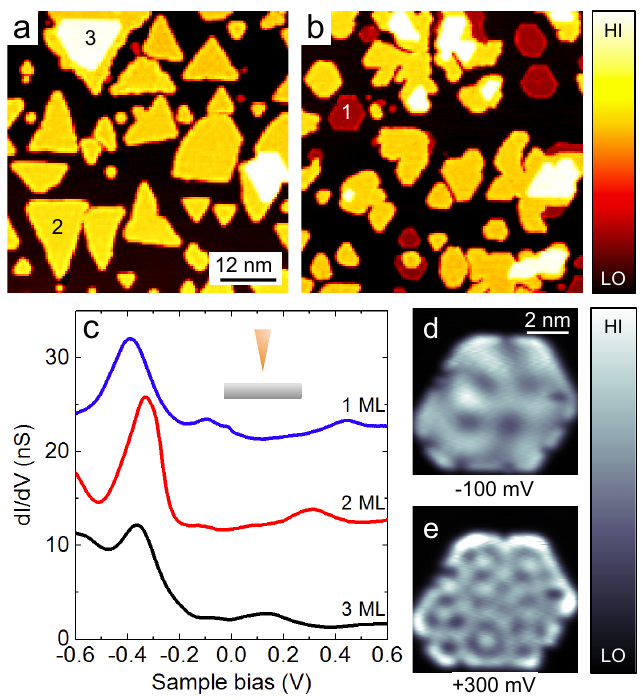}
  \caption{(a) Constant-current image of Co islands grown on (a) Cu(111), (b) ``strained'' Cu(111) (image size: $40\times40$~nm$^2$, sample bias: $80$~mV, tunnel current: $60$~pA). Numbers indicate ML thickness. (c) Typical $dI/dV$ spectra acquired in the center of \textbf{1}, \textbf{2}, and \textbf{3} ML cobalt islands. The 2- and 1-ML spectra are displaced upward by $10$ and $20$ nS, respectively. The feedback loop was opened at $0.5$~V and $1$~nA. The \textbf{2} ML spectrum was acquired on an ``unfaulted'' island \cite{Pietzsch2004}. Inset: Sketch of the tunnel junction. Constant-height $dI/dV$ maps of a Co monolayer acquired at (d) $-100$ mV and (e) $+300$ mV, respectively. The feedback loop was opened in the center of Co at a distance of approximately $500$ pm from tip-surface contact. The dispersive states do not have sufficient weight compared to the $d$ states to be directly detected by tunneling spectroscopy~\cite{Pietzsch2006,Heinrich2009}.
\label{F1}}
\end{figure}

Cobalt deposition results in the formation of triangular-like nanoislands (for details, see supplementary material)~\cite{SM}, which are two atomic layers in height \cite{Diekhoner2003,Pietzsch2004,Rastei2007}. The islands, which are denoted by \textbf{2} in Fig.~\ref{F1}a), display an apparent height of approximately $325\pm 10$ pm and exhibit typical sizes spanning from $10$ to $30$ nm. Approximately $5\%$ of the surface area is occupied by islands that are three layers high (denoted by \textbf{3} in Fig.\ref{F1}a), characterized by an apparent height of $540\pm 10$ pm. To promote the growth of one-layer high islands (referred to as \textbf{1} in Fig.~\ref{F1}b), we introduce strain relaxations in the Cu substrate by either depositing Co at sub-room temperatures or by lowering the annealing temperature of the Cu substrate \cite{Negulyaev2008}. The monolayer islands exhibit variable sizes within the range of $7-20$ nm and have an apparent height of approximately $140\pm 10$ pm. Their morphology is triangular with truncated corners, which signifies that the diffusion barriers involved in their growth process differ from those of bilayer islands \cite{Negulyaev2008}.

The islands exhibit similar electronic properties. In Fig.~\ref{F1}c, we present differential conductance ($dI/dV$) spectra acquired by placing a metal tip at the center of the islands. They were recorded using a lock-in amplifier operating at a frequency of $6.2$ kHz and a modulation of $5$ mV rms. Across all the islands, a prominent feature is observed in their $dI/dV$ spectra, situated at energies of $-0.39$ eV (\textbf{1}), $-0.33$ eV (\textbf{2}), and $-0.36$ eV (\textbf{3}). Based on our \textit{ab initio} calculations (Fig. S1)~\cite{SM}, it is assigned to minority $d_{z^2}$ states, in agreement with previous findings~\cite{Diekhoner2003,Rastei2007}. Their detection in vacuum, however, is possible due to their hybridization with $s-p$ states. Furthermore, similar to the bi- and trilayer islands \cite{Diekhoner2003,Pietzsch2006,Heinrich2009}, the monolayer islands host a free-electron-like surface state. Their confinement within the islands manifests as a standing wave pattern (Fig.~\ref{F1}d-e), exhibiting continuous variations in wavelength across a broad bias range (see Fig. S2 for more wave patterns). The dispersion in the monolayer islands has parabolic behavior with an onset energy of $-0.28\pm0.01$ eV~\cite{SM} and an effective mass of $m^{*}=(0.44\pm0.02)\,m$ ($m$: free electron mass), consistent with that of the bi- and trilayer islands.

The magnetic properties differ for the islands. Their magnetism is explored by attaching a Nc molecule to the tip apex (see Fig. S3) \cite{SM}, and by recording the inelastic signal through the second derivative, $d^2I/dV^2$, of the current $I$ versus bias $V$ (with a lock-in modulation of $500$ $\mu$V rms). Figure~\ref{F2}a displays a series of distance-dependent $d^2I/dV^2$ spectra obtained above a monolayer island with the tip positioned directly above a Co atom at the island's center. Figure~\ref{F2}b presents the corresponding 2D intensity plot. To calibrate the tip-surface distance, we utilized current versus displacement traces above a Co atom (see Fig. S3), setting $z=0$ at the tip-Co contact point. At a distance of $z=100$ pm, the spectrum exhibits a peak and a dip at biases of $+3.9$ and $-3.9$ mV, respectively. These peaks and dips indicate inelastic scattering events, wherein tunneling electrons transfer momentum and energy to the spin states of Nc~\cite{Ormaza2017b}. As the tip approaches, the peak shifts upward, and the dip shifts downward in energy by nearly $5$ mV. At $z\le60$ pm, a second peak (dip) emerges, initially as a high-energy (low-energy) shoulder to the peak (dip), then becoming fully resolved at $z=40$ pm due to its more pronounced energy shift. Intriguingly, in the bilayer and a majority of trilayer islands ($>80\%$), we observe instead a progressive splitting of the peak and dip in the $d^2I/dV^2$ spectrum as the tip approached a Co atom vertically (Fig.~\ref{F2}c and \ref{F2}d; refer to Fig. S6 for the bilayer case). However, a minority of trilayer islands ($<20\%$) exhibit a behavior similar to monolayer islands.

To rationalize these findings, we assign the $z$-axis as the out-of-plane direction of the Co surface, for now assuming the molecular axis of Nc along the $z$-axis~\cite{Ormaza2017b,Czap2019}. We consider the spin Hamiltonian described by Eq.~\ref{eq1}:
\begin{equation}
\hat{H}=D S_z^2+g\mu_\text{B}\hat{B}_\text{ex}\hat{S},
\label{eq1}
\end{equation}
where $\hat{B}_\text{ex}=E_{ex}/g\mu_\text{B}$ represents the exchange field produced by the Co atom on the spin $\hat{S}$ of Nc~\cite{SM}, $\mu_\text{B}$ denotes the Bohr magneton, and $g=1.89$~\cite{Czap2019}. The monolayer spectra are reproduced using a dynamical scattering model with an in-plane exchange field (solid red lines in Fig.~\ref{F2}a), $\hat{B}_\text{ex}=B_\text{ex}\hat{x}$ in Eq.~\ref{eq1} \cite{Ternes2015}. In this configuration, the ground state $\ket{\Psi_0}$ and the two excited states $\ket{\Psi_1}$ and $\ket{\Psi_2}$ consist of a mixing of the $\lvert M\rvert$-spin states (Fig. S4), where $M$ denotes the magnetic quantum number along the $z$-axis. We find a magnetic anisotropy of $D=3.9\pm 0.1$ meV, which is usual for Nc-tips \cite{Ormaza2017a}, and, consistent with expectations, the same exchange energy for both spin excitations (Fig. S5). 

Conversely, the bilayer and trilayer spectra indicate instead that the exchange field has out-of-plane orientation ($\hat{B}_\text{ex}=B_\text{ex}\hat{z}$ in Eq.~\ref{eq1})~\cite{Verlhac2019}. In this configuration, the two excited states, $\ket{\Psi_1}=\ket{M=+1}$ and $\ket{\Psi_2}=\ket{M=-1}$, become Zeeman split (see Fig. S4). The simulated line shapes (solid red lines in Fig.~\ref{F2}c), accurately match the experimental data, but unlike the monolayer islands, a finite lifetime of $\tau\simeq 0.1$ ns is accounted for the excited states (Fig. S7)~\cite{Loth2010a,SM}. The lifetime is nearly $1000$ times longer than that of a magnetic atom on a metal surface~\cite{Khajetoorians2011}. 

\begin{table}
\setlength{\tabcolsep}{3pt}
\begin{tabular}{lcc}
    		&  Mag. moment ($\mu_\text{B}$) & MAE per Co atom (meV)\\
   \hline \hline
   1 ML 	& 1.75  & +0.628  \\
   \hline   
   2 ML  	& 1.76  & -0.208   \\
    \hline 
   3 ML 	& 1.84  & +0.005  \\
   \hline   
\end{tabular}
\caption{DFT-computed magnetic moment of the topmost Co atom and MAE per Co atom for the three islands. More details are given in~\cite{SM}. Positive MAE corresponds to an in-plane orientation of the magnetization.}
\label{T1}
\end{table}

To validate these observations, we conducted Density Functional Theory (DFT) calculations to determine the Magnetic Anisotropy Energy per Co atom (MAE) within the islands (refer to~\cite{SM} for calculation details and an extended presentation of the results). The calculated MAE values (Tab.~\ref{T1}) are consistent with the experimental findings, with one exception being the trilayer case, where the MAE is close to zero. Other effects not accounted for in the simulations, such as the size and shape, therefore likely favor out-of-plane magnetization in the trilayer islands. Additionally, the presence of Cu at the corners of the trilayer islands, as evidenced by the Nc-tip measurements (see Fig.~S8), could also influence the magnetic properties~\cite{Weber1995}. Based on these findings, some variability in the orientation of magnetization for the trilayer islands can be expected, depending on the specifics of the sample preparation.

\begin{figure}
  \includegraphics[width=\columnwidth]{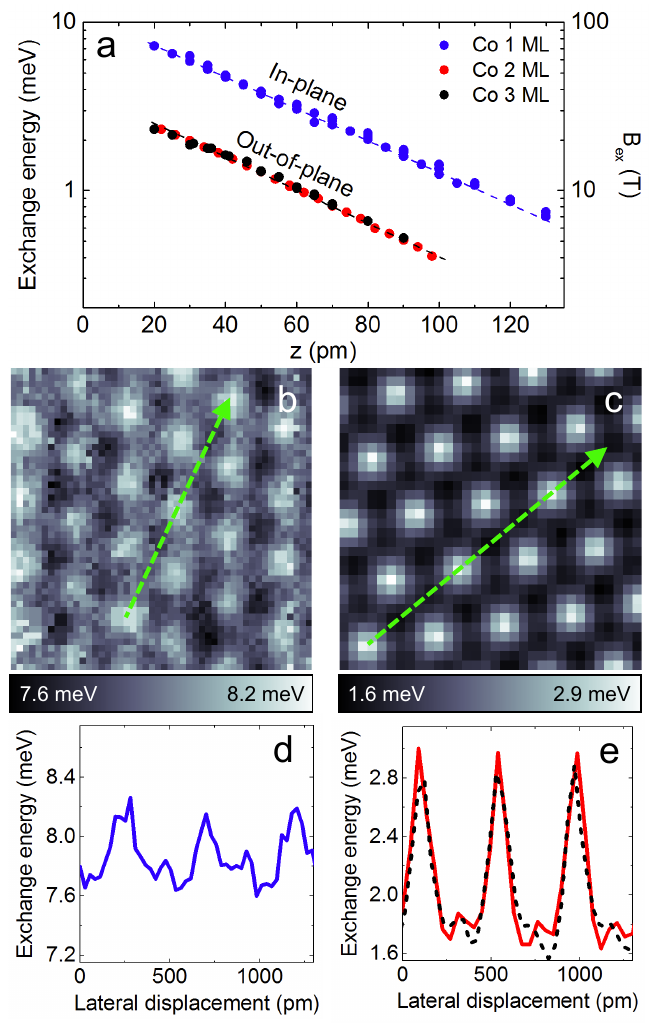}
  \caption{(a) Exchange energy as a function of distance $z$ extracted from the $d^2I/dV^2$ spectra. The dashed lines is the exponential fit described in the text. The exchange field is given on the right $y$-axis. (b,c) Spatial maps of the exchange energy acquired at a fixed distance of $z=20$ pm above a monolayer and a bilayer island, respectively. Lateral (vertical) tip drift between each spectrum acquisition was maintained $<10$ pm ($<5$ pm) by dynamically tracking a local maximum. (d,e) Height profiles of the layers acquired along the dashed lines indicated in (b) and (c). The dashed line in (d) is the height profile of a trilayer.  
\label{F3}}
\end{figure}

\begin{figure}
  \includegraphics[width=\columnwidth]{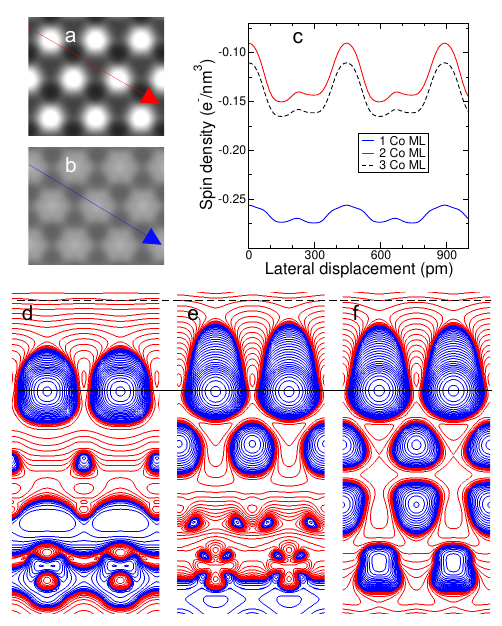}
  \caption{(a)-(b) Computed spin density maps at a fixed distance of $300$ pm above a bilayer and a monolayer, respectively. (c) Height profiles of the layers along the lines indicated in (a) and (b). The dashed line is the height profile of a trilayer. (d)-(f) Section of the spin density in a $(11\bar{2})$ plane passing through the surface Co atom for monolayer, bilayer and trilayer, respectively. Blue (red) isolines indicate positive (negative) values. The black line indicates the position of the surface Co atoms. The height profiles in panel (c) are extracted $300$ pm above the Co atoms (the position is indicated by a dashed black line).
\label{F3theo}}
\end{figure}

\begin{figure}
  \includegraphics[width=1\columnwidth]{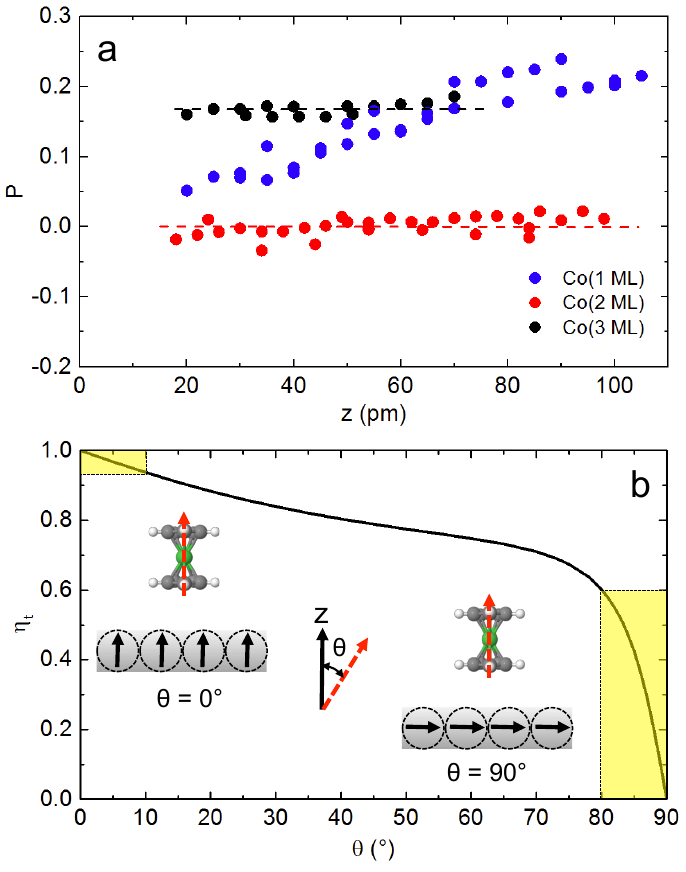}
  \caption{(a) Spin polarization as a function of distance $z$ extracted from the $d^2I/dV^2$ spectra. The dashed lines are a guide to the eye. (b) Computed value of $\eta_t$ as a function of the angle $\theta$. The insets sketch the two extremes where $\theta=0^\circ$ (left) and $\theta=90^\circ$ (right). The light-yellow areas highlight a $10^\circ$ tilt.
\label{F4}}
\end{figure}


The simulated line shapes provide quantitative insights into the exchange energy at specific Nc-Co distances. Above all island types, the exchange energy has an exponential variation $\exp{(-z/\lambda)}$ with a decay length of $\lambda=46\pm3$ pm (Fig.~\ref{F3}a). Notably, the exchange energy above the monolayer island is nearly three times stronger when compared to the bi- and trilayer islands, which exhibit identical exchange energies. The exchange coupling also changed across the surface. To visualize these variations, we acquire a three-dimensional dataset in the form of $d^2I/dV^2 (x,y,V)$ at a fixed distance of $z=20$ pm above a region of interest. We then determine the exchange energy at each lateral tip position by fitting the spectra. The resulting image (Fig.~\ref{F3}b-c) is comparable to the magnetic interaction map obtained for a single atom using STM-ESR \cite{Willke2019}. However, in contrast to STM-ESR, our method allows us to span over multiple surface atoms, as illustrated for monolayer and bilayer islands in Fig.~\ref{F3}b and \ref{F3}c, respectively. These images reveal a magnetic corrugation with a periodicity of $254\pm 6$ pm, corresponding to the in-plane Co-Co distance determined with a metal tip. The corrugation is approximately two times weaker in the monolayer island (Fig.\ref{F3}d) compared to the bi- and trilayer islands (Fig.\ref{F3}e).

This observation is in agreement with the computed spin density maps (Fig.~\ref{F3theo}a and \ref{F3theo}b). As in the experimental exchange energy maps, the corrugation of the spin density is stronger for the bi- and the trilayer islands compared to the monolayer island. The difference can also be observed in the height profiles (Fig.~\ref{F3theo}c), and in the sections of the spin density in a $(11\bar{2})$ plane passing through the Co surface atoms (Fig.~\ref{F3theo}d-f). At $300$ pm (dashed line), the isolines show a stronger corrugation for bi- and trilayers (Fig.~\ref{F3theo}e-f) than for the monolayer (Fig.~\ref{F3theo}d). The origin of this behavior is tracked down to the $d_{z^2}$ orbital. While for the monolayer the orbital has a spin magnetization of 0.28~$\mu_B$, for the bilayer it has a higher spin magnetization of 0.42~$\mu_B$ (0.43~$\mu_B$ for the trilayer). At the same time, the monolayer has a higher average absolute value for the spin density (Fig.~\ref{F3theo}c) in agreement with the stronger exchange energy observed experimentally.

The cobalt islands can also show spin polarization as their magnetization induces a spin imbalance in the tunneling current. This leads to discernible variations in the relative heights of the peaks and dips within the $d^2I/dV^2$ spectrum \cite{Verlhac2019}. For instance, in the case of a trilayer island, the low-energy excitation exhibits a higher dip amplitude at negative bias compared to the peak at positive bias (Fig.~\ref{F2}c; see also Fig.~S7). To quantitatively assess this imbalance, we derive values from our line shape fits in the form of a spin polarization $P=(h_+-h_-)/(h_++h_-)$, where $h_+$ and $h_-$ represent the heights of the peak and dip, respectively. The trilayer islands exhibit a spin polarization of $+0.17$ at the Fermi energy  (Fig.~\ref{F4}a), while the bilayer islands remain non-polarized within the investigated $z$-range.

The spin polarization of the monolayer island shows instead some $z$-dependency (Fig.~\ref{F4}a), changing from $P=+0.2$ at $z=100$ pm to $P=+0.05$ at $z=20$ pm. This variation is attributed to the rotation of the molecular axis relative to the $z$-axis, as predicted by DFT calculations. In vacuum, the tilt angle is close to $10^\circ$, but approaches $0^\circ$ when Nc comes into contact with the surface \cite{Ormaza2017a,Czap2019}. To gain a deeper understanding of this behavior, we introduce the tip polarization ($\eta_t$) and the sample polarization ($\eta_s$), expressing the spin polarization of the tunnel junction as $P=\eta_t\eta_s$. While $\eta_s$ reflects the net polarization in the sample's DOS, $\eta_t$ quantifies the spin momentum transfer during the spin excitation \cite{Loth_2010b}. To compute $\eta_t$, we introduce an angle $\theta$ between $B_\text{ex}$ and the molecular axis and determine the eigenvectors $\ket{\Psi_0}$, $\ket{\Psi_1}$, and $\ket{\Psi_2}$ through Eq. \ref{eq1}. The variation of $\eta_t$ with respect to $\theta$ is presented in Fig.~\ref{F4}b. For small $\theta$, the molecular axis of the Nc-tip aligns nearly parallel to the out-of-plane surface magnetization, and the measured spin polarization closely mirrors the sample polarization ($P\approx\eta_s$). This corresponds to the behavior encountered in the bi- and trilayer islands. Conversely, as $\theta$ approaches $90^\circ$, $\eta_t$ exhibits a pronounced angular dependence. This behavior is characteristic of the monolayer island, where the molecular axis of the Nc-tip is almost perpendicular to the in-plane surface magnetization. For a perfect perpendicular alignment, the tip polarization becomes $\eta_t=0$, resulting in a non-polarized tunnel junction ($P=0$). However, even a relatively small tilt angle of $10^\circ$, corresponding to $\theta=80^\circ$, is sufficient to elevate the tip polarization to $+0.6$. Assuming this tilt angle at $z=100$ pm, we can estimate a sample in-plane polarization of $\eta_s=+0.33$ for the monolayer. 

In summary, our investigation demonstrates the high sensitivity of an Nc-tip to surface magnetism and spin transport. A three-dimensional dataset of spin-excitation spectra allows for the construction of spatial maps of the exchange interaction, which resemble computed spin density maps. The maps offer atomic-level information on the spin orientation of the sample, making them valuable for magnetic imaging, particularly for intricate spin textures. Spatial maps of the surface spin polarization can be constructed by using the amplitude of the inelastic peaks—a facet we intend to explore in forthcoming work. To expand the possibilities of Nc-based magnetometry, future work will involve combining conductance and force measurements in a Nc-functionalized STM-AFM setup.

\begin{acknowledgments}
L.L. gratefully thanks M. Ternes for providing his spin Hamiltonian solver. The Strasbourg authors acknowledge support from the EU’s Horizon 2020 research and innovation programme under the Marie Skłodowska-Curie grant 847471, from the International Center for Frontier Research in Chemistry (Strasbourg), and from the High Performance Computing Center of the University of Strasbourg. Part of the computing resources were funded by the Equipex Equip@Meso project (Programme Investissements d'Avenir) and the CPER Alsacalcul/Big Data. R.B. and N.L. thank financial support from projects RTI2018-097895-B-C44, PID2021-127917NB-I00 funded by MCIN/AEI/10.13039/501100011033, from project QUAN-000021-01 funded by the Gipuzkoa Provincial Council and from project IT-1527-22 funded by the Basque Government. Funded by the European Union. Views and opinions expressed are however those of the author(s) only and do not necessarily reflect those of the European Union. Neither the European Union nor the granting authority can be held responsible for them. 
\end{acknowledgments}



%

\end{document}